# A Natural Copula

Peter B. Lerner[1]

Abstract

Copulas are widely used in financial economics (Brigo 2010) as well as in other areas of applied mathematics. Yet, there is much arbitrariness in their choice. The author proposes "a natural copula" concept, which minimizes Wasserstein distance between distributions in some space, in which both these distributions are embedded. Transport properties and hydrodynamic interpretation are discussed with two examples of distributions of financial significance. A natural copula can be parsimoniously estimated by the methods of linear programming.

## 1. Introduction

Since their invention by Sklar (Sklar, Fonctions de repartition a n dimensions et leur marges 1959), (Sklar, Random functions, joint distributions and copulas 1960), copulas became a valuable instrument in financial mathematics. Possible choices of copulas are numerous: Gaussian, Archimedean, etc. (Durante 2016). In this paper, we propose and motivate a concept of the "natural copula", namely the one, which minimizes Wasserstein distance (Villani 2003) in the space, in which they are mutually embedded. Some arbitrariness remains because the Wasserstein distance can be chosen in multiple ways, but this choice must correspond to the substantive nature of a problem.

We use a VAP (volume-at-price) distribution and an actual limit order book (LOB) to illustrate the concept. Our approximation method allows fast minimization of the Wasserstein distance by the methods of linear programming (Matousek 2007). Furthermore, our result admits an elegant interpretation in hydrodynamic terms, namely that copula corresponds to the flow of incompressible liquid (probability distributions that are normalized to unity) between two distributions viewed as a source and a drain of probability flow.

---

[1] Self

The paper is structured as follows. In section 2 we provide a literature review. In Section 3, the problem is formulated in a relevant language. In Section 4, a parsimonious linear programming method of estimating the copula is given. In Section 5, we provide two examples of natural copulas between 1) the Volume-at-Price distribution of IBM stock, and 2) a trader's LOB of the SDPR500 on an arbitrary trading day. In Section 6, we illustrate a natural copula through a hydrodynamic analogy. In Section 7, a measure of correlation similar to Kendall tau is defined. Finally, we conclude our treatment in Section 8.

2. **Literature review**

The application of copulas is based on a celebrated Sklar (Sklar, Fonctions de repartition a n dimensions et leur marges 1959) (Sklar, Random functions, joint distributions and copulas 1960). In financial mathematics, copulas are used to evaluate the distribution of the random variable X+Y when the marginal distributions for X and Y are known quantities. The distributions for X and Y can be arbitrary and mutually dependent (Cerubini 2016). Also, they are rarely Gaussian.

In particular, because of the additivity of money, the prices of derivative products must be linear functions of the components of the replicating portfolio. In the asset management business, a typical problem is modeling a portfolio of financial instruments, frequently, a credit portfolio with each instrument having its own distribution (Brigo 2010). While the expected payoff of each instrument can be a highly nonlinear function of the parameters, the distribution of the portfolio is a linear combination of the payoffs of its constituent parts.

Finally, in the market microstructure, the distributions of the quotes on the buy and sell side of the LOB are connected by the clearing condition. This requires that the volume of sold securities must be equal to the volume of bought securities and that the orders are filled at the best execution price (Hasbrouck 2007).

We want to present the last application as an example. It has an obvious connection with the transportation problem of Monge (Monge 1781). Gaspar Monge, an XVIII century mathematician, first posed a problem of optimal moving of *remblais* (rubble) into an accurate pile (*deblais*). This problem caused a whirlwind of literature in the XX century, related to programming and optimization, from pioneering work by Kantorovich (Kantorovich 1942) to our days (Brenier 2004).

Our definition of copula as "natural" is such that the distance to carry purchased securities to cover sales, or vice versa, is minimal. Since the 1960s, but mainly in the XXI century, the transport problems acquired an elegant formulation in hydrodynamic terms ( (Brenier 2004) and *op. cit*). We demonstrate the hydrodynamic interpretation of the samples of one-dimensional distributions. The extension of the procedure to a multidimensional case is sometimes non-trivial, but there are algorithmic methods to build (n+1) dimensional copula from the family of n-dimensional copulas (Shemyakin A. 2017).

It would be attractive to think of, for example, 3-dimensional copula as represented by the following form:

$$U_{3D}(x,y,z) = \frac{1}{3!}(U_1(x, U_{23}(y,z)) + U_2(y, U_{13}(x,z)) + U_3(z, U_{12}(x,y)) + perm.) \quad (*)$$

This expression certainly satisfies the requirements of being a copula. However, is not clear to this author that this is the required solution. Furthermore, the estimation of this form according to Section 4 is not straightforward. This problem is currently under investigation.

### 3. Problem design

For our demonstration of the method, we chose the Wasserstein-2 metric as a criterion for the affinity of the distributions. The functional to minimize is thus the following:

$$I = \int (x-y)^2 \, \pi(x,y) dx dy \quad (1)$$

Where

$$\begin{aligned} F_X(x) &= \int_0^x \pi(x,y) dy \\ F_Y(y) &= \int_0^y \pi(x,y) dx \end{aligned} \quad (1a)$$

Are the marginals under constraints:

$$F_X(1) = 1,$$
$$F_Y(1) = 1. \qquad (1b)$$

Here $F_X(x)$ and $F_Y(y)$ are the cumulative distribution functions. To provide a practical solution to the problem, we use parametric approximations of the real financial distributions. A particular parametrization can be chosen in an infinite number of ways, and it is not an integral part of the method. To make my computations feasible for a laptop, I use the following parametrizations for the probability densities and a rather small number (n=4) of Hermite polynomials.

$$\begin{cases} f_X(x) = \left.\sum_{i=1}^{4} a_i H_i(x) e^{-x^2/2} \middle/ \int_0^1 dx \sum_{i=1}^{4} a_i H_i(x) e^{-x^2/2} \right. \\ f_Y(y) = \left.\sum_{i=1}^{4} b_i H_i(y) e^{-y^2/2} \middle/ \int_0^1 dy \sum_{i=1}^{4} b_i H_i(y) e^{-y^2/2} \right. \end{cases} \qquad (2)$$

In the system of Equations (2), $H_i$ are the Hermite polynomials (Abramowitz 1964). We use a Gaussian cutoff factor in (2) to be able to extend integration from the finite sampling domain to the real line $[-\infty, \infty]$.

For the convenience of comparison with the actual volume-at-price distributions, the test functions of Equation (2) were rescaled into the following form:

$$\begin{cases} f_B(x) = \left. V_B \sum_{i=1}^{4} a_i \xi^i H_i\left((x - p_B)/\sigma_p\right) e^{-(x-p_B)^2/2\theta\sigma_p^2} \middle/ \int_0^1 dx \sum_{i=1}^{4} a_i \xi^i H_i\left((x - p)/\sigma_p\right) e^{-(x-p_B)^2/2\theta\sigma_p^2} \right. \\ f_S(y) = \left. V_S \sum_{i=1}^{4} b_i \xi^i H_i\left((y - p_S)/\sigma_p\right) e^{-(y-p_B)^2/2\theta\sigma_p^2} \middle/ \int_0^1 dy \sum_{i=1}^{4} b_i \xi^i H_i\left((y - p)/\sigma_p\right) e^{-(y-p_B)^2/2\theta\sigma_p^2} \right. \end{cases}$$

(2')

In Equations (2'), $V_{B,S}$ are the volume amplitudes for buy and sell distributions, $p_{B,S}$ are the centered prices, $\xi$ is a scaling factor for the polynomial powers, and $\sigma$ – implied width of the price distributions.

## 4. Parsimonious optimization

To minimize the functional (1) under constraints (1a-1b), we use the following approach. This approach is not an integral part of the selection of a natural copula but it is an efficient method of approximating copulas. First, we seek a solution in the form:

$$\pi(x, y) = f_X(x) f_Y(y) \tau(x, y) \tag{3}$$

Where $\tau(x,y) = C + P(x,y)$.[2] Here C is a constant and $P(x,y)$ is a polynomial such that $P(x,0)=0$ and $P(0,y)=0$. The constraints of the system (1b) are satisfied by a specific choice of normalization of the marginals $f_X(x)$ and $f_Y(y)$ of Equation (2). The integration of each monomial in a polynomial $P(x,y)$ provides a certain integral $I_i$. Subsequent integration of each monomial multiplied by a mutual distance, $(x-y)^2$ produces another set of definite integrals, $\tilde{I}_i$. That reduces minimizing the functional (1) to the linear programming problem:

$$\min_{\{c_i\}} \sum_{i=0}^{n} c_i \tilde{I}_i \tag{4}$$

under the following constraint:

$$\sum_{i=0}^{n} c_i I_i \geq 1 \tag{5}$$

Note, that we changed a unity normalization condition for the distribution by an inequality. Because of the properties of a distance, all integrals $\tilde{I}_i$ are non-negative and, consequently, the minimum is always achieved by equality. Otherwise, we could always make an affine scaling of the coefficients so that Equation (5) has the sign of equality but the resulting value of the functional of Equation (4) is lower. With an approximation of each of the marginals by four Hermite functions and four-monomial polynomial $\tau(x,y)$, the Mathematica© optimization takes a few seconds on a PC and, usually, only one or two coefficients in $\tau(x,y)$ are significant.

---

[2] Note that the constant is uniquely determined by the normalization condition.

## 5. Two examples of copulas

To provide examples of the method, we use (1) value-at-price (VAP) distribution of the trades in IBM stock on 05.03.2016 and (2) SPDR500 LOB distribution of one trader on or around 03.01.2014. Both distributions were smoothed down with MA(2) process before applying a parametric approximation. We estimated real distributions according to Equations (2'). After that, they were projected into $[0,1]\times[0,1]$ squares and the optimization of the copula according to Section 4 was obtained. The parameters for two examples (four one-dimensional distributions) are given in Tables 1 and 2.

[Place Tables 1 and 2 here]

Empirical distributions and their estimates are plotted in Figs. 1 and 2.

[Place Figures 1 and 2 here]

## 6. Hydrodynamic interpretation of the "Natural Copula"

The connection of the problems of optimal transport with hydrodynamics evolved from the 1960s work by V. I. Arnold (Arnold 1966) and was resuscitated in the 2000s through the work of Benamou and Brenier (see (Brenier 2004) and *op*. *cit*.). [3] Approximately speaking, the Wasserstein distance between two shapes/distributions is minimized by the flow of incompressible fluid, which obeys Euler equations. Heretofore, we can assume that one marginal distribution is a hydrodynamic source and the other marginal distribution is a drain for an incompressible fluid. Incompressibility preserves the normalization of the distributions of probability flows. The properties of a natural copula are thus related to the global properties of probability flow.

---
[3] For the modern bibliography, one can consult (Liu 2016) and (Chizat 2016).

We illustrate this interpretation by plotting the vector field associated with the derivatives of the copula function in Fig. 3. We interpret these derivatives as velocities:

$$\begin{cases} v_x = \frac{\partial V}{\partial y} \\ v_y = -\frac{\partial V}{\partial x} \end{cases} \quad (6)$$

Where V(x,y) is a potential function of the flow, which we identify with our copula. Then, as usual, we can define the following characteristics:

$$\begin{cases} \Gamma = \oint v_x \, dx + v_y dy \\ \Phi = \oint v_x \, dy - v_y dx \end{cases} \quad (7)$$

which are called circulation and flux in hydrodynamics. We choose the contour of integration along the border of the sampling area. Because of the square integration region [0,1]×[0,1], the y-integrals over parts of the integration contour parallel to the x-axis and the x-integrals over contour intervals parallel to the y-axis are identically zero (see Fig. ?). So, only one component of flow velocity gives a contribution to the contour integral:

$$\begin{cases} \Gamma = \int_{\{0,0\}}^{\{1,0\}} v_x \, dx + \int_{\{1,0\}}^{\{1,1\}} v_y \, dy + \int_{\{1,1\}}^{\{0,1\}} v_x \, dx + \int_{\{1,1\}}^{\{0,1\}} v_y \, dy \\ \Phi = -\int_{\{0,0\}}^{\{1,0\}} v_y \, dx + \int_{\{1,0\}}^{\{1,1\}} v_x \, dy - \int_{\{1,1\}}^{\{0,1\}} v_y \, dx + \int_{\{1,1\}}^{\{0,1\}} v_x \, dy \end{cases} \quad (8)$$

The results of our calculation in arbitrary units are provided in Table 2.

[Place Table 3]

We tentatively identify circulation with the number of shares changing hands in the case of optimal unwinding of the LOB and flux – with the market imbalance.

## 7. Measure of correlation

The measure of correlation, which is usually invoked in connection to copulas is the Kendall copula tau parameter (Shemyakin A. 2017). A natural extension of the concept of correlation to our method of estimation is the following expression:

$$C_T = \int f_X(x) f_Y(y)(\tau^2(x,y) - 1) dx dy \tag{9}$$

For the factorizable distributions, the expression of Equation (9) is zero by construction. The estimates of our two model distributions are given in Table 4. For both selected distributions, the $C_T$ is on the order of 20%. The above correlation is independent of our copula approach and is only the feature of the preferred method of estimation, which we outlined in Section 5.

[Place Table 3]

## 8. Conclusion

This paper demonstrates that it is possible to define a "natural" copula, which minimizes Wasserstein distance between distributions in a space, in which they are both embedded. Its estimation does not involve any arbitrariness other than a particular embedding space of the marginals, which usually appears naturally in the problem and the index of the Wasserstein distance.

We propose a parsimonious method of estimation of a natural copula through the linear programming algorithm. This method can be applied independently of any parametric form of the copula approximation.

The natural copula has a hydrodynamic interpretation. Namely, if we consider one distribution a drain, and another—a sink, the flow of probability between them obeys Euler equations. The incompressibility of the liquid corresponds to the preservation of the normalization for the distributions.

Table 1. Parametric approximation for two examples of Section 4. The marginals are parametrized as follows: $\tilde{f}_B(x) = Max[0, \tilde{f}_B(x)], \tilde{f}_S(x) = Max[0, \tilde{f}_S(x)]$ and

$$\begin{cases} f_B(x) = \dfrac{V_B \sum_{i=1}^{4} a_i \xi^i H_i ((x-p_B)/\sigma_p) e^{-(x-p_B)^2/2\theta\sigma_p^2}}{\int_0^1 dx \sum_{i=1}^{4} a_i \xi^i H_i ((x-p)/\sigma_p) e^{-(x-p_B)^2/2\theta\sigma_p^2}} \\ f_S(y) = \dfrac{V_S \sum_{i=1}^{4} b_i \xi^i H_i ((y-p_S)/\sigma_p) e^{-(y-p_B)^2/2\theta\sigma_p^2}}{\int_0^1 dy \sum_{i=1}^{4} b_i \xi^i H_i ((y-p)/\sigma_p) e^{-(y-p_B)^2/2\theta\sigma_p^2}} \end{cases}$$

| Sample | $V_B$ | $V_S$ | $\sigma_p$, $ | $\xi$ | $p_B$, $ | $p_S$, $ | $\theta$ |
|---|---|---|---|---|---|---|---|
| IBM050316 | 3.5 | 6.0 | 0.0471 | 3.558 | 13.374 | 13.561 | 1 |
| SDPR500 | 14.0 | 16.0 | 10.561 | 1.698 | 173.164 | 174.116 | 2 |

Table 2. Parametrization of the function $f(x,y) = \dfrac{1+Ax^2y}{2}$ of Equation (3).

| Sample | A |
|---|---|
| IBM | 27.66 |
| SDPR500 | 9.78 |

Table 3. Hydrodynamic flux $\Phi$ and circulation $\Gamma$ (Equation (6)) for the sample copulas in arbitrary units. Turnover is significant, while the imbalance is small.

| Sample | $\Phi$ | $\Gamma$ |
|---|---|---|
| IBM | 25.97 | -0.285 |
| SDPR500 | 61.03 | 0.045 |

Table 4. The measure of correlation $C_T = \int f_X(x) f_Y(y) (\tau^2(x,y) - 1) dx dy$, Equation (8) for the normalized copulas.

| Sample | $C_T$ |
|---|---|
| IBM | 0.2011 |
| SDPR500 | 0.1861 |

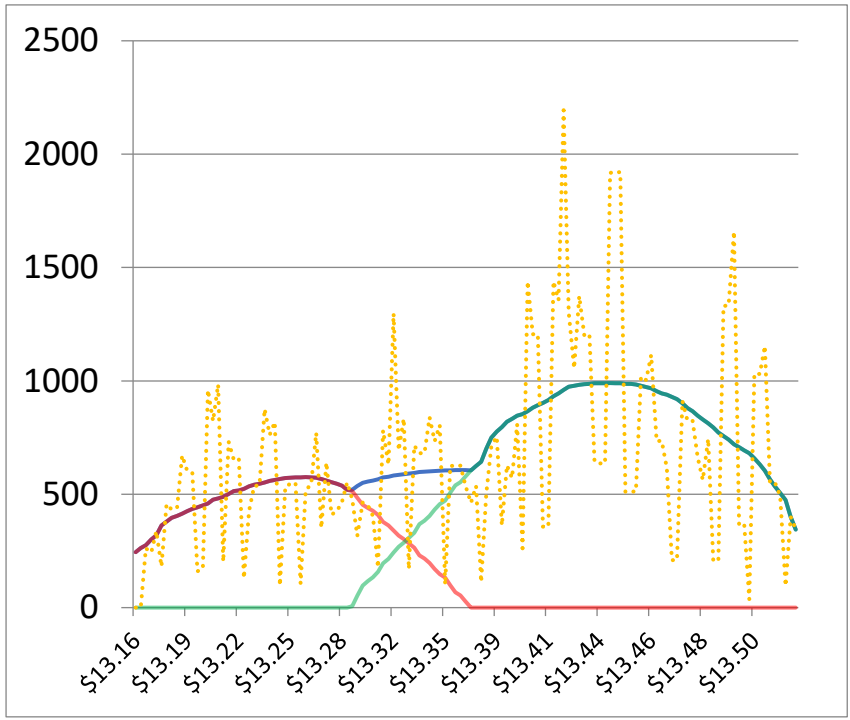

Fig. 1 Volume-at-Price distribution (yellow dot) for the IBM stock on May 03, 2016. The solid red line indicates our approximant of offers, the solid green line is the approximant of bids and the blue curve is the sum of the total.

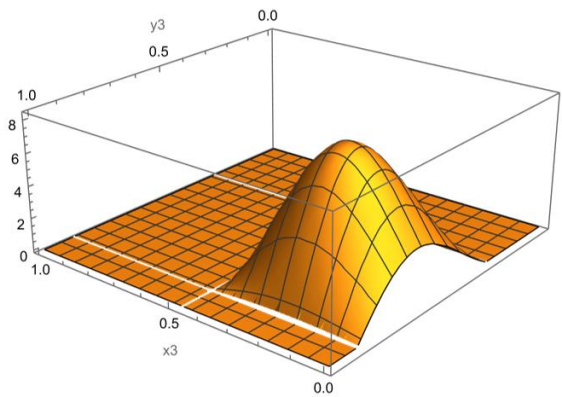

A)

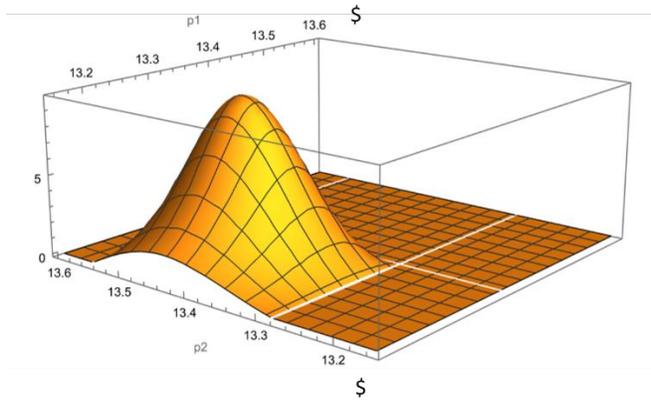

B)

Fig. 2. The density of the natural copula of the IBM VaP distribution in A) normalized coordinates, B) original dollar prices. The vertical scale is arbitrary.

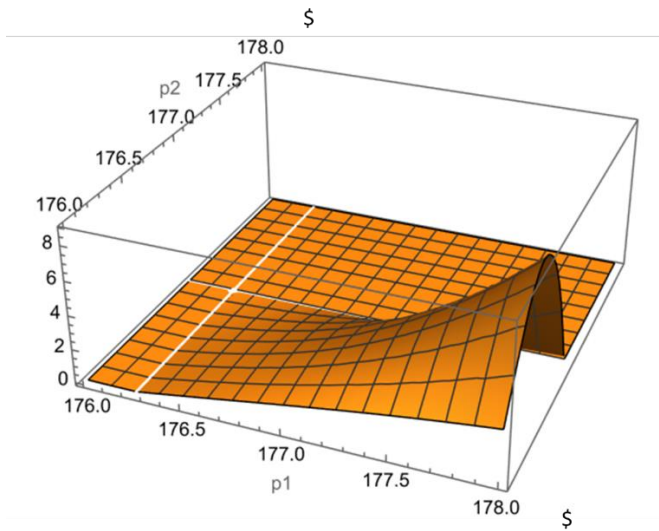

Fig. 3. The density of a natural copula of a sample LOB of SPDR500 on an average trading day.

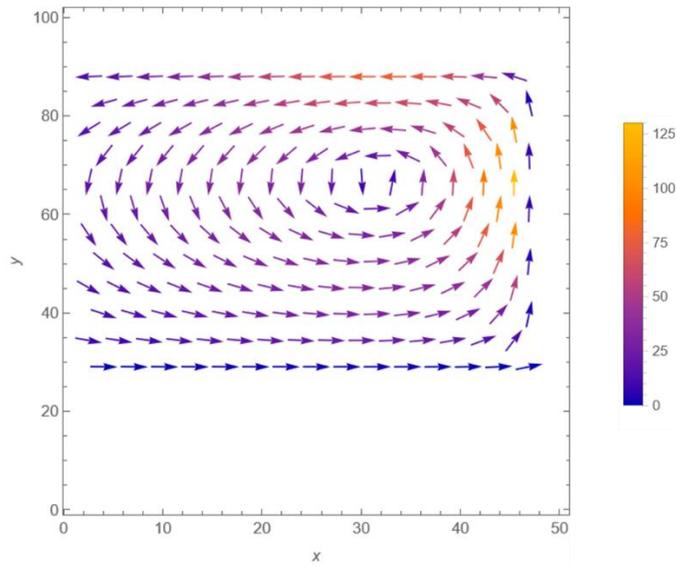

A)

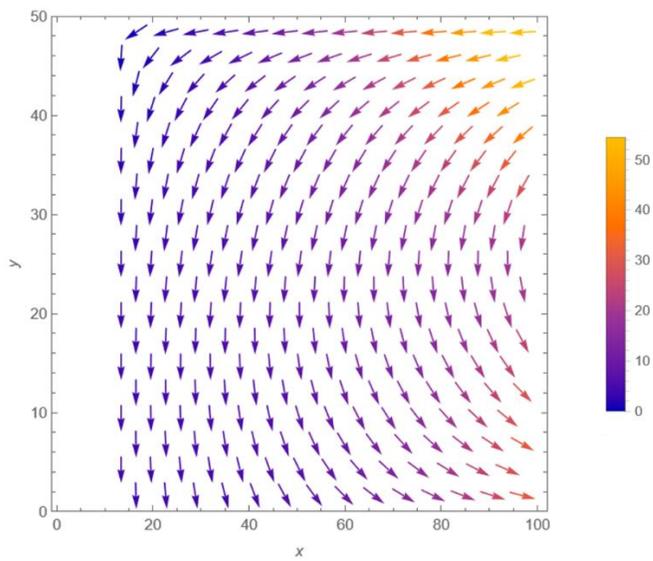

B)

Fig. 4. Vector fields of hydrodynamic velocities (Equation (6) of the main text) for the copula distributions of Figs. 2 and 3. The coordinates x and y are the percentage scales of the embedding box [0,1]×[0,1].